\documentclass{sf2a-conf2011}
\usepackage{graphicx}
\usepackage{hyperref}
\usepackage[]{natbib}  

\def\BibTeX{{\rm B\kern-.05em{\sc i\kern-.025em b}\kern-.08em
    T\kern-.1667em\lower.7ex\hbox{E}\kern-.125emX}}
\bibpunct{(}{)}{;}{a}{}{,}  
\newcommand{\kms}{km\,s$^{-1}$}
\newcommand{\lsim}{~\rlap{$<$}{\lower 1.0ex\hbox{$\sim$}}}

\newcommand{\HI}{H\,{\sc {i}}~}

\newcommand{\Msold}{M$_{\odot}$\,yr$^{-1}$}

\newcommand{\Teff}{T$_{\rm eff}$}

\begin{document}

\TitreGlobal{SF2A 2011}


\title{An H\,{\sc {i}} 21-cm line survey of evolved stars}

\runningtitle{An H\,{\sc {i}} 21-cm line survey of evolved stars}

\author{E. G\'erard}\address{GEPI, Observatoire de Paris}

\author{T. Le~Bertre}\address{LERMA, Observatoire de Paris}

\author{Y. Libert}\address{IRAM}



\setcounter{page}{1}

\index{G\'erard, E.}
\index{Le~Bertre, T.}
\index{Libert, Y.}


\maketitle


\begin{abstract}
The \HI line at 21 cm is a tracer of circumstellar matter around
AGB stars, and especially of the matter located at large distances
(0.1-1 pc) from the central stars. It can give unique
information on the kinematics and on the physical conditions
in the outer parts of circumstellar shells and in the regions
where stellar matter is injected into the interstellar medium.
However this tracer has not been much used up to now, due to
the difficulty of separating the genuine circumstellar emission
from the interstellar one.

With the Nan\c cay Radiotelescope we are carrying out a survey
of the \HI emission in a large sample of evolved stars.
We report on recent progresses of this long term
programme, with emphasis on S-type stars.
\end{abstract}



\section{Introduction}
  Low- to intermediate-mass stars, at the end of their evolution, 
become red giants. In this phase they may undergo mass loss 
at a very large rate ($>$ 10$^{-8}$ \Msold), even so large that 
it has a decisive effect on their late evolution \citep{o99}. 
Observations show that the rate at which this phenomenon 
develops varies highly from source to source, so that the balance of mass 
loss as a function of the initial conditions (mass, metallicity, etc.) and of 
the stage of evolution is presently not well understood.

The \HI line at 21 cm is potentially well suited to determine the history 
of mass loss because hydrogen is the dominant element in AGB outflows 
and because atomic hydrogen should be protected from photoionization by 
the surrounding interstellar medium (ISM). However, the dominant species in 
the atmospheres of AGB stars is expected to be atomic hydrogen only in 
relatively "warm" red giants with \Teff $>$ 2500 K, and by contrast 
it should be molecular hydrogen in "cool" red giants with \Teff $<$ 2500 K 
\citep{gh}. Nevertheless molecular hydrogen 
should be ultimately photodissociated by the interstellar radiation 
field in the external parts of circumstellar shells \citep{mj}. 
The distance at which this happens, is expected to depend mainly on the square 
root of the mass loss rate, but this needs to be proven. Also molecular 
hydrogen might survive at larger distance, if the outflows are clumpy.
Atomic hydrogen should thus be also a useful tracer of the physical 
conditions in the outer parts of circumstellar shells and in the regions 
where stellar matter is injected into the ISM. 
Therefore, \HI spectro-imagery of circumstellar environments is expected 
to bring a wealth of information on the relations between stars and the ISM.

\section{\HI surveys}

However the detection of red giants in the \HI line at 21 cm happens to be 
difficult partly due to the weakness of the signal, and even more to the 
competing emission by the ISM on the same lines of sight. Pionneering efforts 
in the 1980's led to the detection of only two sources, $\alpha$ Ori and Mira 
\citep{bk87,bk88}.  
This topic was abandoned until we readdressed it in 2001, after the renovation 
of the Nan\c cay Radio Telescope (NRT). In a first survey we  
detected \HI in about 20 red giants \citep{glb06}. 
The high sensitivity to low surface brightness emission, a good spatial 
resolution in right ascension (4$'$ at 21 cm) and a well-adapted observing 
technique (position-switch with increasing east-west offset) 
contributed to this success. 

The majority of the sources that were detected have a mass loss rate 
of only a few 10$^{-7}$ \Msold. Also the majority have a warm central star 
with \Teff $>$ 2500 K. These two results are not independent because, for red 
giants, there is an inverse correlation between \Teff ~and \.M, with warm 
sources undergoing a relatively low mass loss rate, as compared to cool 
sources (with  \Teff \lsim ~2500 K) that may have rates as large as 10$^{-5}$ 
to 10$^{-4}$\,\Msold. The line profiles of our detected \HI sources 
are generally narrower than the CO line profiles, 
implying a slowing down of stellar outflows by surrounding material. 
The emissions are extended, indicating shell sizes on the order of 1 pc, 
and suggesting the possibility of tracing the history of mass loss over a few 
10$^5$\,years (see following section). 

\begin{figure}[ht!]
 \centering
 \includegraphics[width=0.33\textwidth,clip,angle=270]{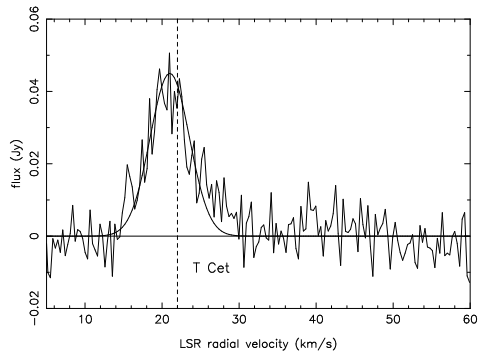}  
  \caption{\HI spectrum of T Cet. The vertical dashed line marks the radial 
velocity determined from circumstellar CO lines by \citet{rso}. 
The thin line is a gaussian fit to the \HI profile.}
  \label{lb:fig1}
\end{figure}

In collaboration with L.D. Matthews, the most promising sources are then 
imaged with the Very Large Array \citep[VLA; e.g.][]{mlglbr,mlglbj}. 
These images allow us to study in more detail the geometry of the 
circumstellar shells which tend to exhibit signatures of ISM interaction 
due to the motion of the central stars with respect to the local ISM 
\citep{vgsm,llbg}. 

The observation of \HI in cool sources (\Teff \lsim ~2500 K) is more difficult 
because atomic hydrogen is not expected to peak on the central star, 
but rather 
to be found only at some distance as a daughter species of molecular hydrogen. 
For nearby sources we therefore need to cover large areas on the sky in order 
to detect the putative \HI emission (that may or may not be present 
depending on the history of mass loss, clumping,...). Although more difficult 
to study, we cannot ignore these sources which may in fact dominate the 
replenishment of the interstellar medium by stellar matter \citep{s94}. 

It thus appears necessary to survey systematically evolved stars for their 
\HI emission. We have already performed systematic observations of a sample 
of 60 carbon stars \citep{glbl}, and presently study a sample 
of 20 S-type stars, i.e. AGB stars with a photospheric C/O abundance ratio 
close to 1 or slightly less, mainly selected from the 
work of \citet{rso}. These authors have obtained CO radio 
observations that allow them to derive accurate radial velocities which, 
for us, are a useful guide to identify the circumstellar \HI lines. 

\begin{figure}[ht!]
 \centering
 \includegraphics[width=0.3\textwidth,clip,angle=270]{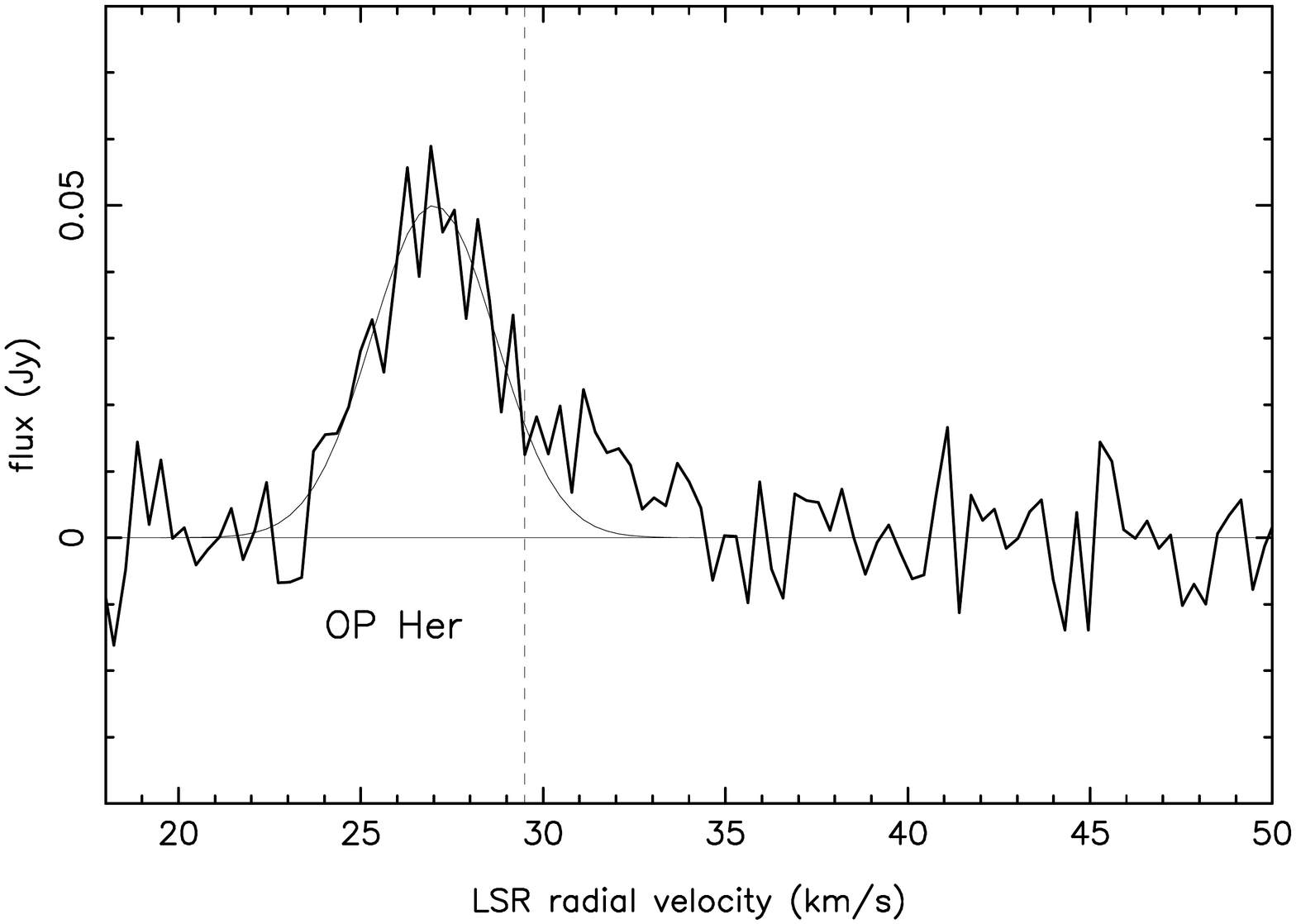}\hspace*{1cm}
 \includegraphics[width=0.3\textwidth,clip,angle=270]{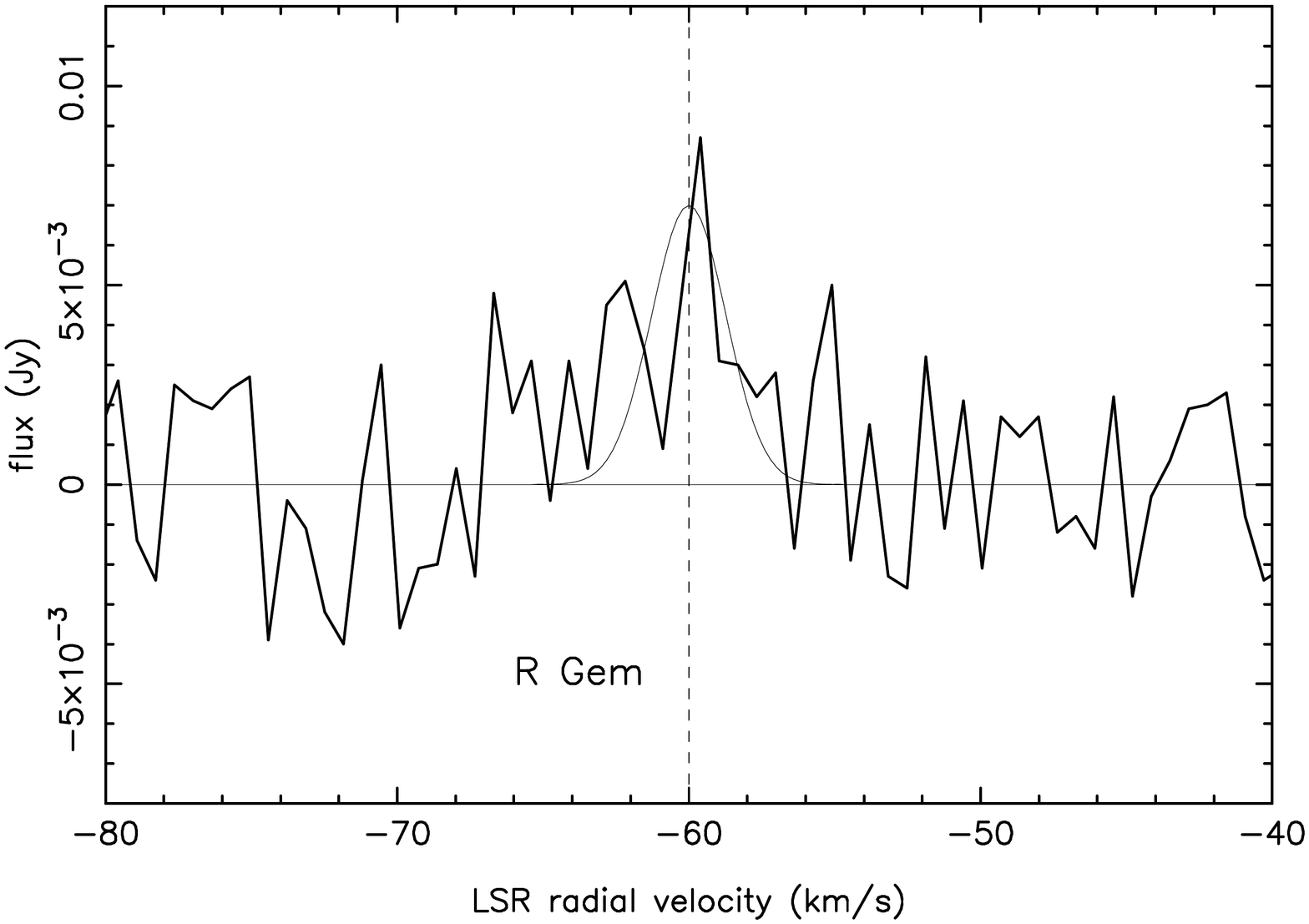} 
  \caption{{\bf Left:} \HI spectrum of OP Her. The vertical dashed line marks 
the stellar radial velocity from \citet{f05}.  
 {\bf Right:} \HI spectrum of R Gem. The vertical dashed line marks the radial 
velocity determined from circumstellar CO lines by \citet{rso}. 
The thin lines are gaussian fits to the \HI profiles.}
  \label{lb:fig2}
\end{figure}

In Fig.~\ref{lb:fig1}, we show the \HI profile of T Cet, a S-type semi-regular 
variable (SRb), with a mass loss rate of 0.4 10$^{-7}$ \Msold ~\citep[for a 
distance of 240\,pc,][]{rso}. The \HI line is slighly offset 
in velocity ($\sim$\,1\,\kms) with respect to the CO lines. As usual, this 
offset is towards the zero-velocity of the local standard-of-rest (LSR) 
frame which suggests a dragging 
of the circumstellar envelope by the ambient ISM \citep{glb06}. 
In  Fig.~\ref{lb:fig2} (left), we show the spectrum of OP Her, an irregular 
variable (Lb) of type S for which no radio CO detection has been reported. 
\citet{f05} give a LSR radial velocity of 
29.5~\kms, whereas we find an \HI line centered at 27~\kms.

T Cet and OP\,Her are rather warm sources at high galactic latitudes, 
and we easily detect their \HI emissions. On the other hand the S-type 
Mira variable R Gem is barely detected in \HI (Fig.~\ref{lb:fig2} right), 
although it is a strong CO emitter, with a substantial mass loss rate 
\citep[4.4 10$^{-7}$ \Msold, for a distance of 710 pc,][]{rso}. 
As its radial velocity is large ($-60$ \kms), confusion remains weak, and 
we could obtain a marginal detection of $\sim$ 7 mJy. We presently do not 
fully understand the reason for the weakness of the \HI emission of many 
Miras. Hydrogen could be mainly in molecular form, and preserved from 
photo-dissociation in the outer layers of their circumstellar shells 
(self-shielding in high density regions ?). Alternatively the duration of 
the mass loss phenomenon could be too short to allow for the presence 
of a detectable quantity of atomic hydrogen. It is noteworthy that, 
in their IRAS survey of evolved stars, \citet{ypk} 
found that Mira variables tend to have been losing mass for a shorter period 
than semi-regular variables, and have smaller infrared diameters. 
This questions the often cited argument that semi-regular variables might  
evolve into Mira variables. 

\section{Detailed studies of individual sources}

In our first \HI survey \citep{glb06} we found that, in 
general, the emissions of nearby sources are extended, indicating shell sizes 
on the order of 1 pc, and suggesting the possibility of tracing the history 
of mass loss over a few 10$^5$ years. Also, the \HI emissions are 
sometimes spatially shifted w.r.t. to the central stars 
in a direction that is often opposite to that of the proper motion. 
Using the VLA, a "head-tail" morphology has been found in several cases 
\citep{mr,mlglbr,mlglbj}. Thus it appears that, 
in \HI at 21 cm, we are probing a region that is shaped by the motion of 
the star relative to the ISM. Even though the NRT has a large beam 
(4$'$ in right ascension and 22$'$ in declination at 21 cm) compared 
to the VLA ($\sim$\,1$'$ in the D-configuration),  
the circumstellar environments of many sources can be resolved. In the extreme 
case of Mira, a 2-degree long tail has been discovered in the far ultra-violet 
by the satellite GALEX \citep{msn}. In that case the \HI emission 
extends clearly beyond the VLA primary beam, and observations with a large 
size single dish antenna are needed to detect H\,{\sc {i}} in the tail 
\citep{mlglbr}. It is in such regions that stellar matter is expected 
to be injected into the ISM, and observations at high spectral resolution of 
the \HI line at 21 cm allow us to constrain the kinematics in this kind of 
fascinating, but barely studied, environment. 

\begin{figure}[ht!]
 \centering
 \includegraphics[width=0.29\textwidth,clip,angle=0]{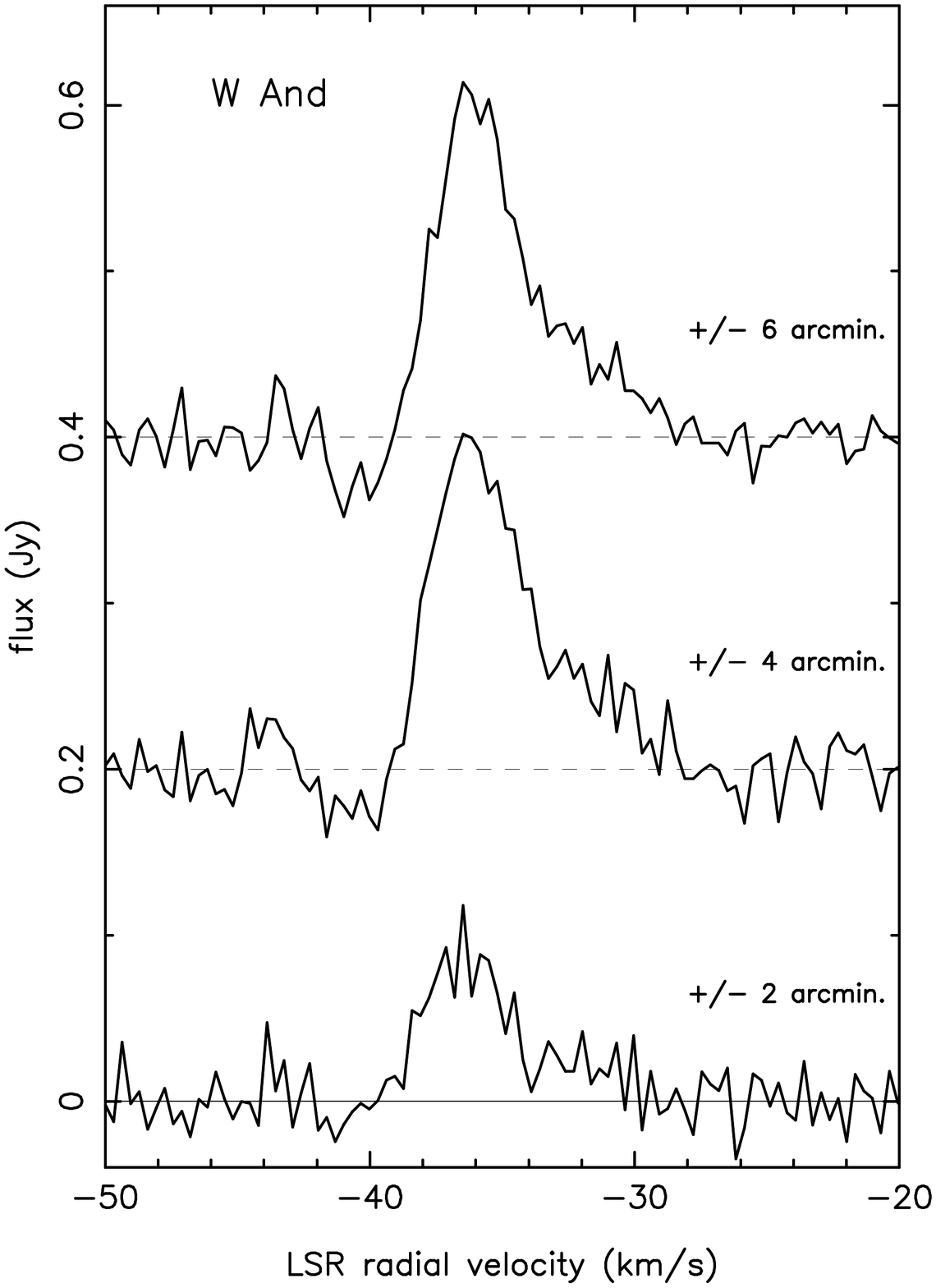}\hspace*{2cm}
 \includegraphics[width=0.29\textwidth,clip,angle=0]{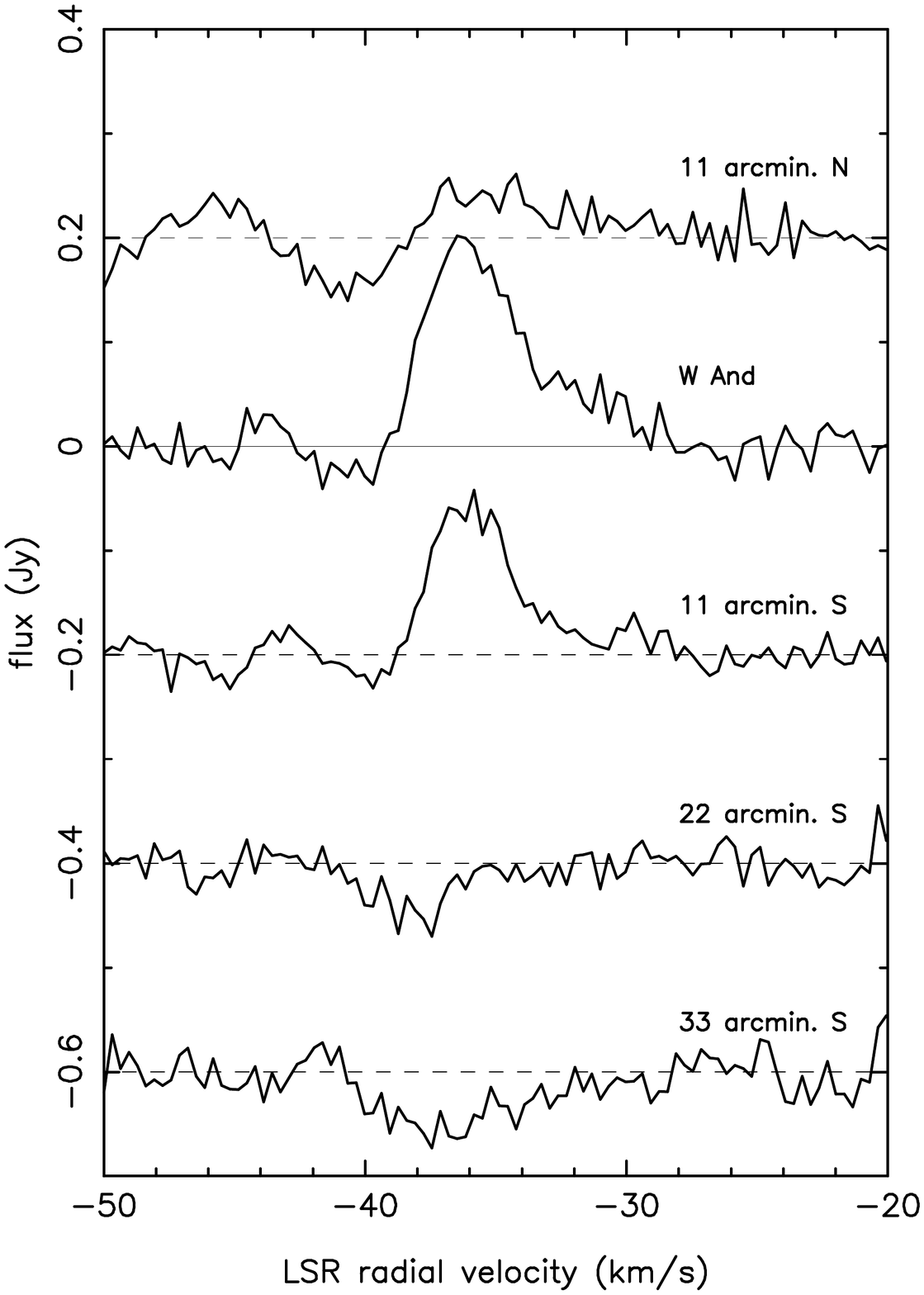}       
  \caption{{\bf Left:} W And spectra obtained in the position-switch mode 
with the star centered (``on'') and the off-positions taken in the east-west 
direction at $\pm 2'$, $\pm 4'$, and $\pm 6'$. 
For clarity the spectra have been successively shifted by 0.2 Jy. 
 {\bf Right:} position-switch spectra ($\pm 4'$) with the central beam 
placed at $+11'$ (half a beam, north), on source (as on the left panel), 
$-11'$ (half a beam, south), $-22'$ (one beam, south), and $-33'$ (one and 
half beam, south).}
  \label{author1:fig3}
\end{figure}

In Fig.~\ref{author1:fig3}, we show the spectra obtained on and around W And, 
a Mira variable of type S, with a mass loss rate of 1.7\,10$^{-7}$ \Msold 
~\citep[for a distance of 280 pc,][]{rso}. 
In the left panel, the position-switch 
spectra obtained at $\pm 4'$ and at $\pm 6'$ are almost indistinguishable. 
Thus the source does not appear to be extended in right ascension (diameter 
size $\leq 4'$). On the other hand the maximum of \HI emission is clearly 
offset south ($\sim 5'$) because the fluxes obtained on the star position 
and at $11'$ south are almost the same (Fig.~\ref{author1:fig3}, right panel). 

There is also a practical reason for mapping individual sources. 
The confusion with 
interstellar emission can be so intricate that a large size map around the 
target is needed in order to isolate the genuine circumstellar emission. 
For example an extended mapping around RS\,Cnc, a S-type semi-regular variable 
(SRc), by \citet{llbgm}, revealed a structure of $\sim$\,18$'$ 
that we could separate from the underlying ISM emission.

\section{Conclusion and future work}
The 21-cm \HI emission from evolved stars brings unique information on the 
kinematics and the physical conditions in the external regions of 
circumstellar shells. However, it is weak, especially in cool sources where 
molecular hydrogen may dominate, 
and its observation often suffers from confusion by the ISM emission. 
More work is needed to understand the formation of \HI lines in circumstellar 
environments, and to assess the potential contribution of \HI studies to 
the physics of circumstellar shells around evolved stars. 

The Square Kilometre Array (SKA) would be ideal to study these weak and 
extended sources. In the \HI line at 21 cm it will be able to image nearby 
circumstellar environments at all scales needed to probe 
the history of mass loss and the interaction of stellar outflows with the ISM. 
Nevertheless, using the observational strategy that we have 
developed with the NRT we can already begin to explore this field. 
The EVLA offers also the promising possibility to study these sources at the 
sub-arcminute scale, as well as will soon the SKA precursors 
(ATA, MeerKAT, etc.) that are under construction.

\begin{acknowledgements}
{\bf Acknowledgements\\}
We are grateful to Drs L.D. Matthews and J.M. Winters for fruitful 
discussions. We thank the PCMI for financial support. The Nan\c{c}ay Radio 
Observatory is the Unit\'e scientifique de Nan\c{c}ay of 
the Observatoire de Paris, associated as Unit\'e de Service et de Recherche 
(USR) No. B704 to the French Centre National de la Recherche Scientifique 
(CNRS). The Nan\c{c}ay Observatory also gratefully acknowledges the financial 
support of the Conseil R\'egional de la R\'egion Centre in France.
\end{acknowledgements}


%
\end{document}